\documentclass[12pt]{article}
\usepackage{graphicx}
\usepackage{cite}
\newcommand{\dissum}[2]{\displaystyle \sum_{#1}^{#2}}

\newcommand{\fnd}[2]{\frac{\textstyle #1}{\textstyle #2}}
\newcommand{\xrm}[1]{{\textstyle \mbox{\rm #1}}}
\newcommand{\bm}[1]{\mbox{\boldmath $#1$}}
\newcommand{\abs}[1]{\left| #1\right|}
\newcommand{\ket}[1]{\mbox{$\left| #1\right\rangle$}}
\newcommand{\bracket}[2]{\mbox{$\left\langle #1\left| #2\right.\right
\rangle$}}
\newcommand{\braket}[3]{\mbox{$\left\langle #1\left|
#2\right| #3\right\rangle$}}
\newcommand{\bra}[1]{\mbox{$\left\langle #1\right|$}}

\newcommand{\babar}
{{\it B}$\!${\footnotesize\it A}$\!${\it B}$\!${\footnotesize\it A$\!$R}}

\begin{document} \baselineskip .7cm
\title{\bf Explaining the \bm{D^{\ast}_{sJ}(2317)^{+}}}
\author{
Eef van Beveren\\
{\normalsize\it Centro de F\'{\i}sica Te\'{o}rica}\\
{\normalsize\it Departamento de F\'{\i}sica, Universidade de Coimbra}\\
{\normalsize\it P-3004-516 Coimbra, Portugal}\\
{\small eef@teor.fis.uc.pt}\\ [.3cm]
\and
George Rupp\\
{\normalsize\it Centro de F\'{\i}sica das Interac\c{c}\~{o}es Fundamentais}\\
{\normalsize\it Instituto Superior T\'{e}cnico, Edif\'{\i}cio Ci\^{e}ncia}\\
{\normalsize\it P-1049-001 Lisboa Codex, Portugal}\\
{\small george@ajax.ist.utl.pt}\\ [.3cm]
{\small PACS number(s): 11.55.Bq, 12.39.Pn, 13.75.Lb, 14.40.Lb} \\[1.0cm]
{\normalsize Talk given at}\\
{\normalsize\it The 25th annual Montreal-Rochester-Syracuse-Toronto
Conference on High-Energy Physics}\\ {\normalsize\bf Joefest}\\
{\normalsize in the honor of the 65th birthday of Joseph Schechter}\\
{\normalsize May 13 - 15, 2003, Syracuse (NY)} \\[0.3cm]
{\small hep-ph/0306155}
}

\maketitle

\begin{abstract}
The recently observed $D^{\ast}_{sJ}(2317)^{+}$ meson is explained as a
scalar $c\bar{s}$ system which appears as a bound state pole,
below threshold, in the $DK$ scattering amplitude.
The standard $c\bar{s}$ charmed scalar $D_{s0}$
is found at about 2.9 GeV, with a width of some 150 MeV.

For the scattering length of $DK$ in $S$ wave we predict $5\pm 1$ GeV$^{-1}$.
\end{abstract}

\section{Introduction}

The $D^{\ast}_{sJ}(2317)^{+}$ state has, since its discovery
\cite{HEPEX0304021} by the \babar\ collaboration (see also
Refs.~\cite{HEPEX0305017,HEPEX0305100}),
raised several interesting questions on its nature
\cite{HEPPH0305012,HEPPH0305025,HEPPH0305035,
HEPPH0305038,HEPPH0305049,HEPPH0305060,
HEPPH0305122,HEPPH0305140,HEPPH0305209,
HEPPH0305213,HEPPH0305292,HEPPH0305315}.
However, the light-scalar-meson hypothesis from unitarization
\cite{ZPC30p615} has made important progress
with the discovery of the $D^{\ast}_{sJ}(2317)^{+}$ state
(most likely $J=0$).
Not often it happens that in experiment a state is discovered
which seems difficult to handle within other theories
\cite{PRD43p1679,PRD48p4370,PRD49p409,PLB388p154,PRD64p114004},
but can be explained by the unitarization hypothesis
without extra ingredients.

Of course, this happened before with the light scalar mesons, which can easily
be handled within a unitarized model for all non-exotic mesons
\cite{HEPPH0201006}, without making any distinction between the heavier scalar
mesons and the light ones.
But seventeen years after the publication of these results \cite{ZPC30p615},
their implications are still not fully understood, nor applied in data
analysis.

In the specific case of the $D^{\ast}_{sJ}(2317)^{+}$ state,
just by applying the model parameters established twenty years ago
\cite{PRD27p1527}, the unitarization scheme results in a bound state
below the $DK$ threshold,
for a $c\bar{s}$ confinement spectrum with ground state at 2.545 GeV,
almost 200 MeV above the $DK$ threshold.
This achievement is entirely similar to the {\it extra} \/states obtained by
the same model for the light scalars.
No additional interactions are necessary to explain the scalar mesons.

\section{Unitarization}

In the unitarization scheme, we describe meson-meson scattering
by the wave equation

\begin{equation}
\psi_{f}\left(\vec{r}\;\right)\; =\;
\left( E-H_{f}\right)^{-1}\; V_{t}\;
\left( E-H_{c}\right)^{-1}\; V_{t}\;\psi_{f}\left(\vec{r}\;\right)
\;\;\; .
\label{scatteqn}
\end{equation}
The operators $H_{f}$ and $H_{c}$ respectively describe
the dynamics of two free mesons and of a confined quark-antiquark system.
The latter is supposed to generate an infinity of confinement states.
Transitions from confined quarks to free mesons, mediated by
quark-antiquark-pair creation, is described by $V_{t}$.

By comparison of Eq.~(\ref{scatteqn}) with the usual expressions for
scattering wave equations, we conclude that the generalized
potential $V$ for meson-meson scattering is given by

\begin{equation}
V\; =\; V_{t}\;\left( E-H_{c}\right)^{-1}\; V_{t}
\;\;\; .
\label{generalpot}
\end{equation}

With the definitions given in Appendix~\ref{Tmtrx},
in particular formulae (\ref{Opmodel}), (\ref{calJcalH})
and Table~\ref{masses}, we obtain for
the partial-wave scattering phase shift generated by wave equation
(\ref{scatteqn}) the result

\begin{equation}
\xrm{cotg}\left(\delta_{\ell}(p)\right)\; =\;
\fnd{\pi\lambda^{2}\mu p\;
\dissum{n=0}{\infty}
\fnd{{\cal J}^{\ast}_{n\ell}(p)\;{\cal N}_{n\ell}(p)}
{E(p)-E_{n\ell_{c}}}\; -\; 1}
{\pi\lambda^{2}\mu p\;
\dissum{n=0}{\infty}
\fnd{{\cal J}^{\ast}_{n\ell}(p)\;{\cal J}_{n\ell}(p)}
{E(p)-E_{n\ell_{c}}}}
\;\;\; .
\label{partialpshift}
\end{equation}
The expression (\ref{partialpshift}), which is an exact solution of
Eq.~(\ref{scatteqn}), is very compact and elegant.
But the summations over the radial excitations of the confined system
do not converge fast. Hence, for the numerical calculation of scattering
quantities it is not suitable.
A more practical method has been described in Ref.\cite{CPC27p377}.
Nevertheless, because of its close contact with the physical situation,
expression (\ref{partialpshift}) is extremely comprehensible,
and useful as a starting point for making approximations.

For example, when we are studying a limited range of energies,
we may restrict the summation to only the few terms near in energy,
characterized by $E_{n\ell_{c}}$ in Eq.~(\ref{partialpshift}),
and approximate the remaining terms by a simple function of $p$,
the relative meson-meson linear momentum.
Notice, however, that the resulting expression is only of the type of
a Breit-Wigner (BW) expansion when the strength of the transition,
characterized by $\lambda$, is small.
For larger values of $\lambda$, expression (\ref{partialpshift}) behaves
very differently from BW expansions.

Another, in practice very useful approximation is inspired by the radial
dependence of the transition potential $V_{t}$, which is depicted in
Ref.~\cite{ZPC21p291} for meson-meson scattering in the cases
$J^{P}=0^{-}$, $J^{P}=1^{-}$, and $J^{P}=0^{+}$,
We find there that $V_{t}$ is peaked at short distances.
If we, moreover, take into account that eigenfunctions of the confinement
Hamiltonian $H_{c}$ must also be of short range, then we may just define a
transition radius $a$ and approximate the spatial integrals of formula
(\ref{calJcalH}) by choosing a spherical delta shell $U(r)=\delta (r-a)$.
In this case we end up with the expression

\begin{equation}
\xrm{cotg}\left(\delta_{\ell}(p)\right)\;\approx\;
\fnd{2a^{4}\lambda^{2}\mu p\;
j_{\ell}(pa)\; n_{\ell}(pa)\;
\dissum{n=0}{\infty}
\fnd{\abs{{\cal F}_{n\ell_{c}}(a)}^{2}}
{E(p)-E_{n\ell_{c}}}\; -\; 1}
{2a^{4}\lambda^{2}\mu p\;
j^{2}_{\ell}(pa)\;
\dissum{n=0}{\infty}
\fnd{\abs{{\cal F}_{n\ell_{c}}(a)}^{2}}
{E(p)-E_{n\ell_{c}}}}
\;\;\; .
\label{partialpshifta}
\end{equation}
This way we have pulled the $p$ dependence outside the infinite summation
over the radial confinement spectrum, which
makes it much easier to handle truncations, as now the rest term does not
depend on $p$ and may thus be chosen constant.
Our recent detailed analyses of scalar mesons have actually been
performed in the approximation of formula~(\ref{partialpshifta}).
\clearpage
\mbox{}

\begin{figure}[htbp]
\centerline{\scalebox{0.7}{\includegraphics{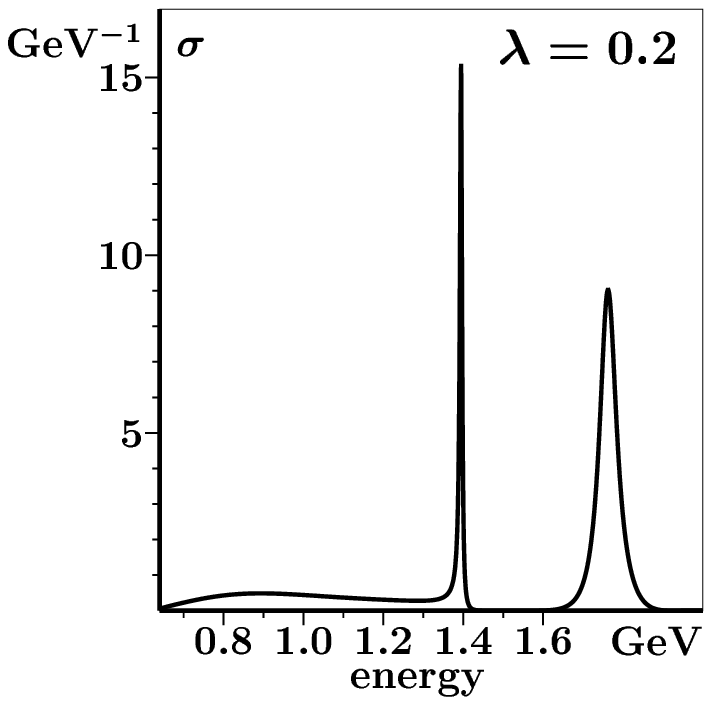}}
\scalebox{0.7}{\includegraphics{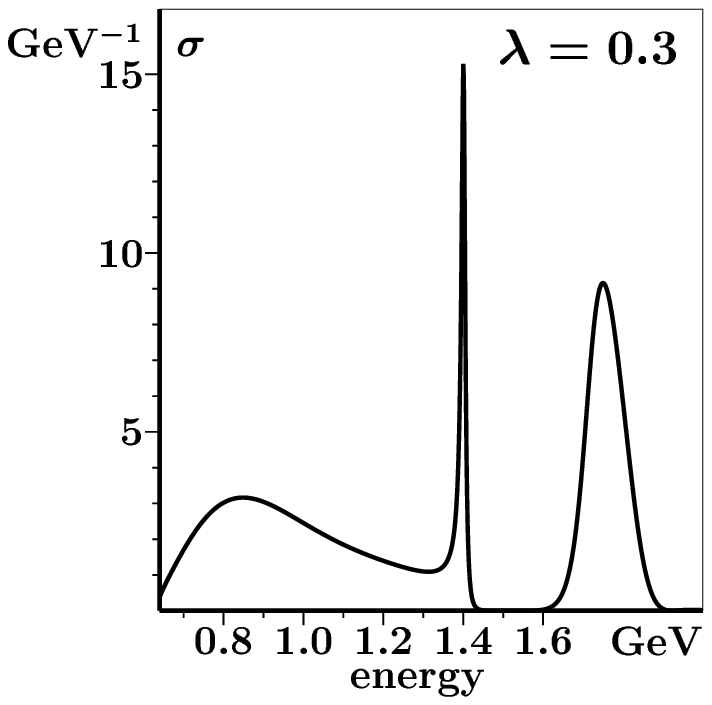}}
\scalebox{0.7}{\includegraphics{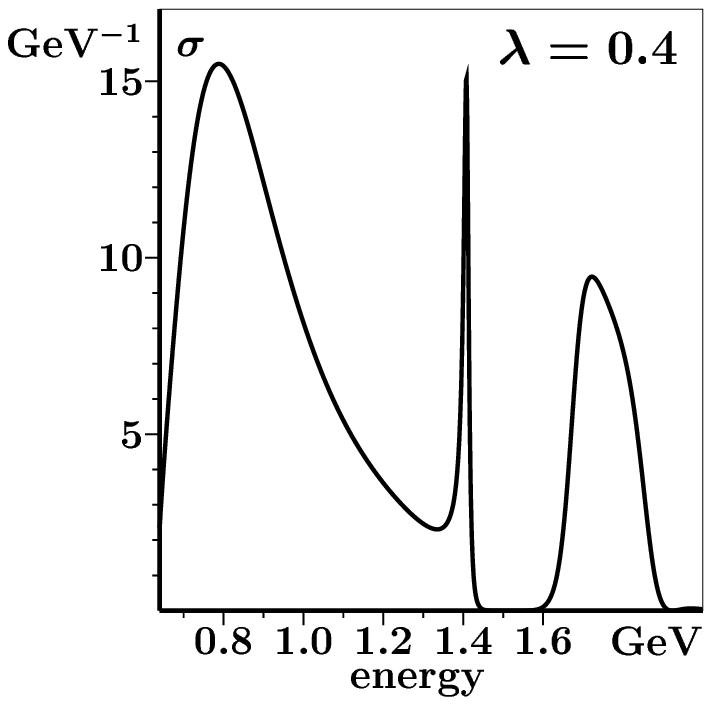}}}
\centerline{\scalebox{0.7}{\includegraphics{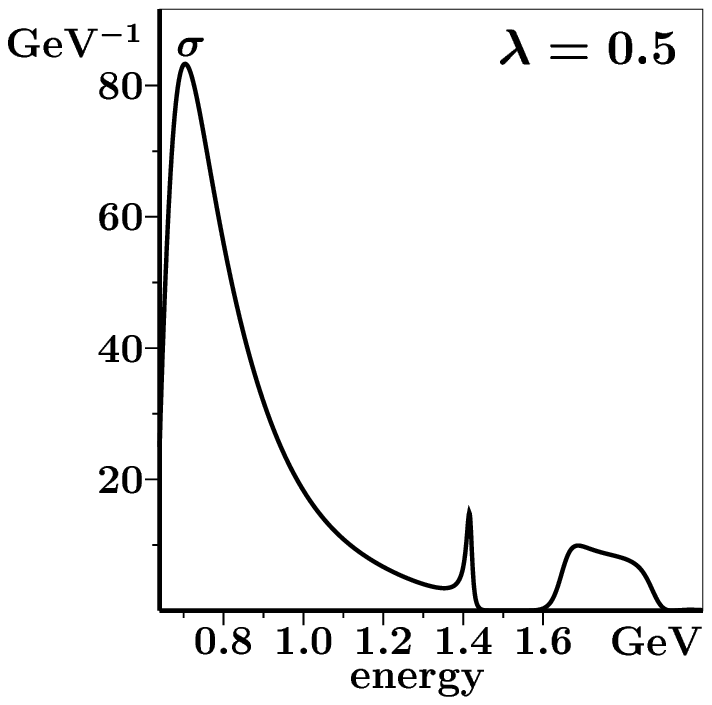}}
\scalebox{0.7}{\includegraphics{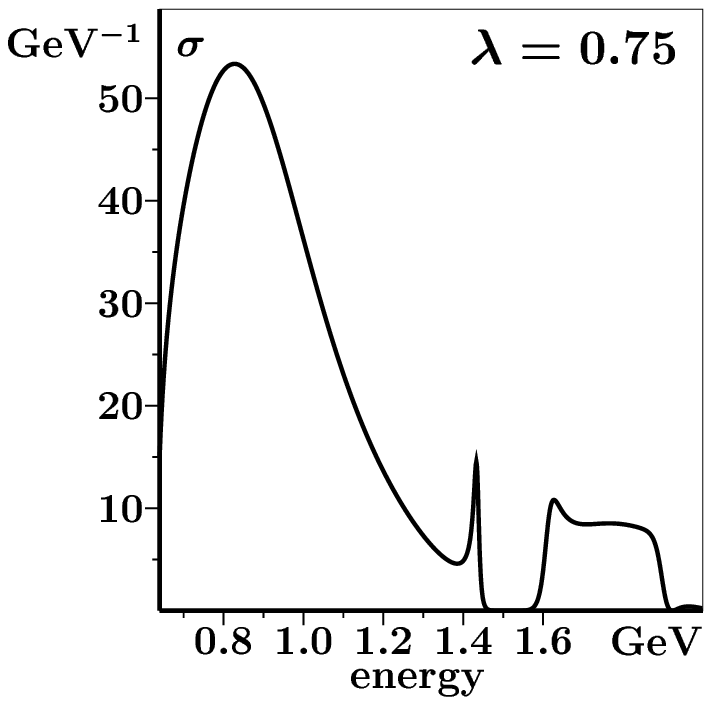}}
\scalebox{0.7}{\includegraphics{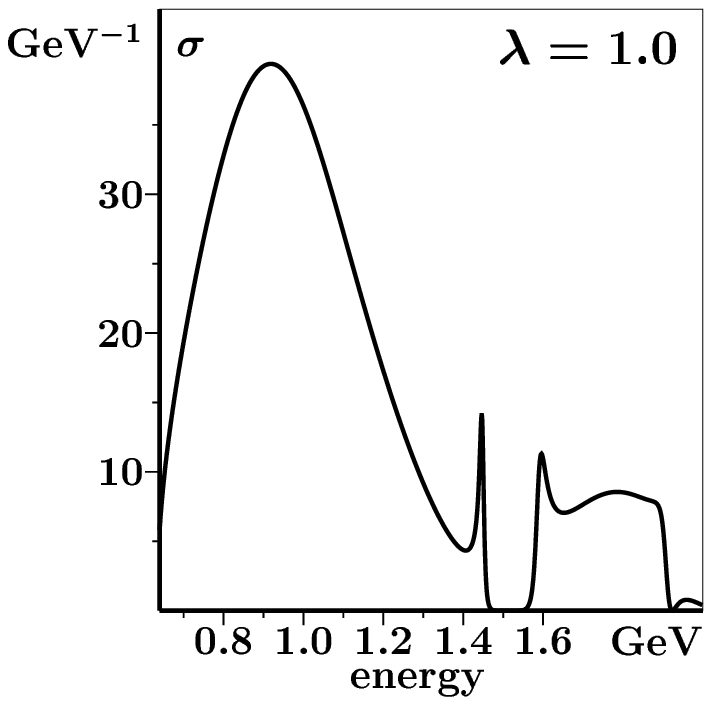}}}
\normalsize
\caption{Theoretical cross section $\sigma$ as a function of the strength
of the transition potential, characterized by $\lambda$,
for $K\pi$ elastic $S$-wave scattering, coupled to a confinement spectrum
with ground state at 1.389 GeV and level spacing of 380 MeV
(model parameters taken from Ref.~\cite{PRD27p1527}).}
\label{cs}
\end{figure}

In Fig.~(\ref{cs}) we show how the model's cross section varies with
$\lambda$ in the approximation given in Eq.~(\ref{partialpshifta}),
using harmonic-oscillator confinement.
The first two plots ($\lambda =$ 0.2 and 0.3) show separate peaks,
not very different from a sum of BW resonances.
For larger values of $\lambda$, the cross section gets more and more
distorted, certainly not suited for a description in terms of separate
or overlapping BW resonances.
The physical value for the strength of the transition potential
comes in our model at around $\lambda =0.75$, for which value we find a broad
structure peaking at some 830 MeV, representing the recently confirmed
\cite{PRL89p121801,ZPC30p615} $K_{0}^{\ast}$(800) (or $\kappa$) resonance.
Note that the latter structure is already showing up for
smaller coupling, and is clearly not related to the ground state
of the $J^{P}=0^{+}$ $s\bar{d}$ confinement spectrum,
at 1.389 GeV for the model parameters from Ref.~\cite{PRD27p1527}.

The plots for $\lambda =0.75$ and $\lambda =1.0$
in Fig.~\ref{cs} show structures in the scattering cross section
that are certainly not well approximated by sums over BW resonances.
Nevertheless, expression~(\ref{partialpshift}) has several features in common
with BW resonances.

A BW resonance is characterized by a singularity in the lower-half of the
complex-energy plane (second Riemann sheet for single-channel scattering).
The real part $E_{0}$ and the imaginary part $\Gamma /2$ of the pole
are about equal to the central resonance position and half the width
of the resonance, respectively.

In our model, as we have the disposal of an analytic expression, given in
Eq.~(\ref{partialpshift}), we can determine the pole positions exactly.
For small values of $\lambda$, the relation between the central peak and
width of a structure in the cross section can rather well be read from
the real and imaginary part of the pole.
However, for larger values of $\lambda$, poles of the scattering amplitude
are not always close to the corresponding peaking structure in the cross
section.

Nevertheless, poles are very useful, because they constitute a clean tool to
relate phenomena of different hadronic systems
\cite{ZPC30p615,AIPCP660p353,PRD51p2353,NUCLTH0212008,HEPPH0301049}.
For example, only through the hypothetical movements of the poles
in the complex-energy plane we can show that the $f_{0}$(600), $f_{0}$(980),
$K_{0}^{\ast}$(800) \cite{PRL89p121801}
and $a_{0}$(980) are manifestations of the same phenomenon, but for different
flavors \cite{EPJC22p493}.

\section{The scalar mesons}

The scalar mesons represent the most noteworthy result \cite{ZPC30p615}
of the unitarized model given in formula~(\ref{scatteqn}).
In the recent past, scalar mesons have been the subject of many investigations,
in a variety of approaches.
Strong and electroweak interaction properties of scalar mesons
were studied in Refs.~\cite{PLB454p365,PRD65p114011,HEPPH0304031},
quark and glue contents of the isoscalars $f_{0}$
in Refs.~\cite{HEPPH0201299,HEPPH0210400,HEPPH0212117,PLB559p49,NUCLEX0302007,
HEPPH0302137,NUCLEX0302013,HEPPH0302133,HEPPH0305043,HEPPH0303248,HEPPH0210028,
HEPPH0208123,NUCLEX0302014,HEPPH0302059,HEPPH0304193,Montanet2003},
the $\kappa$(800) in Refs.~\cite{HEPPH0211026,EPJC22p493,PLB454p365,QCD02p0,
HEPPH0212383},
the isovectors $a_{0}$ in Ref.~\cite{PLB462p14,JPG28p2783},
dynamically generated scalar mesons
in Ref.~\cite{HEPPH0203149},
analogy with the Higgs boson
in Refs.~\cite{HEPPH0305083,HEPPH0304075,HEPPH0201171},
effects of scalar mesons in decay processes and coupling to meson-meson
scattering
in Refs.~\cite{PRD60p034002,DBlack:thesis,PLB536p59},
$a_{0}$/$f_{0}$ mixing in Ref.~\cite{HEPPH0202069,HEPPH0210431,
HEPPH0303223,HEPPH0301126},
mixing of light and heavy scalar mesons
in Refs.~\cite{HEPPH0305296,HEPPH0306031},
and properties of scalar mesons at finite temperatures in
Ref.~\cite{NUCLTH0204055}.

Our understanding of the scalar mesons is based on studying the corresponding
pole positions, and their movements as a function of the coupling $\lambda$.
A pole which stems from a confinement state moves towards the energy
eigenvalue of that state when $\lambda$ is changed to smaller values.
However, we observe that besides the poles stemming directly from the
confinement spectrum, there exist poles in the scattering amplitude which
disappear into the background for vanishing $\lambda$.
These somehow originate in the bulk of the sum over the radial excitations in
expression~(\ref{partialpshift}), and are not related to any state in
particular.

Moreover, we observe that the lightest $\ell=1$ $q\bar{q}$ scalar mesons can
only reside somewhere between 1.3 and 1.5 GeV in the confinement spectrum,
with the model parameters of Ref.~\cite{PRD27p1527}.
The latter have been adjusted to bottomonium, charmonium, the light vector
states, and to $P$-wave meson-meson scattering data, whenever
available in the early eighties.

Finally, we observe that the extra poles in the scattering amplitude
come close enough to the real-energy axis only in the case of $S$-wave
scattering. There, they show up somewhere close to the lowest threshold,
for energies below the ground states of the confinement spectrum.

For the physical value of the model parameter $\lambda$,
a specific value has been selected by Nature.
But in our model we can vary $\lambda$.
This way, we can find out where poles are moving
for smaller or larger values of the coupling.

\section{The \bm{D^{\ast}_{sJ}(2317)^{+}} state}

For $DK$ elastic $S$-wave scattering, we adopt the same expression
that has been used for $K\pi$ as described in Ref.~\cite{EPJC22p493}.
We only have to substitute the confinement masses of $d\bar{s}$
by those for $c\bar{s}$, thereby using the parameters of
Ref.~\cite{PRD27p1527}.
For the ground state of the $J^{P}=0^{+}$ $c\bar{s}$ confinement spectrum,
the mass comes out at 2.545 GeV.
The first radial excitation is then 380 MeV (the model's level spacing) higher
in mass.  The summation over the higher radial excitations in
Eq.~(\ref{partialpshifta}) we approximate by a constant, normalized to 1.
The relative couplings of the ground state and the first radial
excitation to the $DK$ channel are 1.0 and 0.2, respectively \cite{EPJC22p493}.
Thus we obtain
\begin{equation}
\sum_{n=0}^{\infty}\;
\fnd{\abs{{\cal F}_{n,1}(a)}^{2}}
{E(p)-E_{n,1}}\;
\longrightarrow\;
\fnd{1.0}{E(p)-2.545}\; +\; \fnd{0.2}{E(p)-2.925}\; -\; 1
\;\;\;\;\;\xrm{GeV}^{\: 2}
\;\;\; .
\label{DKSwave}
\end{equation}

\begin{figure}[htbp]
\centerline{\scalebox{0.8}{\includegraphics{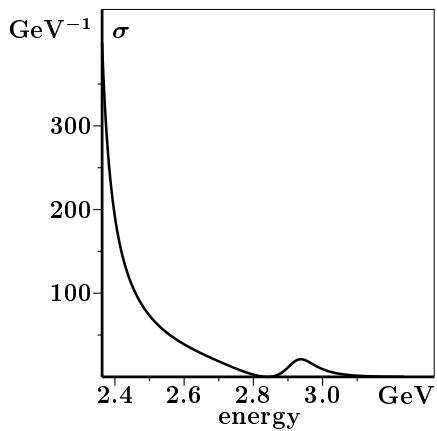}}}
\normalsize
\caption{Theoretical cross section $\sigma$ for $DK$ S-wave elastic
scattering
(model parameters taken from \cite{PRD27p1527}).}
\label{Ds}
\end{figure}

\noindent
For expression~(\ref{partialpshifta}) with the substitution~(\ref{DKSwave}),
and furthermore for the physical value of $\lambda$ (meaning the value
fitting $K\pi$ elastic $S$-wave scattering from threshold up to 1.6 GeV
\cite{EPJC22p493}), we depict the resulting $DK$ elastic $S$-wave cross
section in Fig.~\ref{Ds}.
The corresponding pole positions in the complex-energy plane
for the $D^{\ast}_{s0}$ ground state and first radial excitation
are shown in Ref.~\cite{HEPPH0305035}.
These are found at 2.28 GeV and ($2.78-i\, 0.093$) GeV, respectively .

Notice from Fig.~\ref{Ds} that in the region around 2.55 GeV,
where quark models predicted $J^{P}=0^{+}$ $c\bar{s}$ states
\cite{PRD43p1679,PRD64p114004},
we find no structure in the cross section,
nor poles in the scattering amplitude.
On the contrary, employing confinement \em plus \em \/unitarization,
we find structures in the cross section some 250 MeV below and 400 MeV
above that value, which
explains the observation of the (probable) $D^{\ast}_{s0}$ ground state below
the $DK$ threshold, and predicts for the first radial excitation a resonance
somewhere between 2.8 and 3.0 GeV, with a width of some 100--200 MeV.
Based on Fig.~\ref{Ds}, we may also predict the scattering length for
$DK$ elastic $S$-wave scattering.
Keeping in mind that the $J^{P}=0^{+}$ $c\bar{s}$ ground state comes out
a little bit too low in mass with our value of $\lambda$,
we estimate the theoretical error by varying $\lambda$.
For a conservative choice, we obtain

\begin{equation}
\xrm{scattering length $DK$ in $S$ wave }\; =\; 5\pm 1
\;\;\;\xrm{GeV}^{-1}
\;\;\; .
\label{scatl}
\end{equation}

\section{Conclusions}

The two most important aspects of strong interactions for hadrons are
confinement and OZI-allowed real or virtual decay, which effectively describe
gluonic interactions and quark-pair creation/annihilation, respectively. These
are the only ingredients entering the model defined in Eq.~(\ref{scatteqn}),
which, with a fixed small set of parameters, can reasonably describe hadronic
phenomena ranging in energy from the light scalars up to bottomonium.

Here, we have discussed the model's results for the $J^{P}=0^{+}$ $c\bar{s}$
states coupled to the $DK$ channel. For the $DK$ $S$-wave scattering length we
predict the value $5\pm 1$ GeV$^{-1}$.

\appendix

\section{The \bm{T} matrix for meson-meson scattering}
\label{Tmtrx}

First, we introduce the operators $H_{c}$, describing
the confinement dynamics in the interaction region, $H_{f}$, representing the
dynamics of the scattered particles at large distances, and $V_{t}$, which
stands for the transitions between these two sectors.
For an arbitrary spherically symmetric confinement potential $V_{c}$,
we define these operators in configuration space by
\begin{equation}
\begin{array}{c}
H_{c}\; =\; -\fnd{\nabla^{2}_{r}}{2\mu_{c}}\; +\; m_{q}\; +\; m_{\bar{q}}\;
+\; V_{c}(r)
\;\;\;\;\; ,\;\;\;\;\;
H_{f}\; =\; -\fnd{\nabla^{2}_{r}}{2\mu}\; +\; M_{1}\; +\; M_{2}
\;\;\;\;\; ,\;\;\;\;\; \\[3mm]
\xrm{and}\;\;\;\;\;
V_{t}\; =\;\lambda\; U\left( r\right)
\;\;\;\; .
\end{array}
\label{Opmodel}
\end{equation}
The various mass parameters of Eq.~(\ref{Opmodel}) are defined in
Table~\ref{masses}.
The transition potential $V_{t}$, which provides the communication between
the confined channel and the scattering channel,
is in agreement with the potentials \cite{ZPC21p291} that may describe
the breaking of the color string.

\begin{table}[hbpt]
\begin{center}
\begin{tabular}{|c||l|}
\hline\hline
symbol & definition \\
\hline
$m_{q}$ $\left( m_{\bar{q}}\right)$ & constituent (anti-)quark mass \\
$\mu_{c}$ & reduced mass in confinement channel \\
$M_{1,2}$ & meson masses \\
$\mu$ & reduced mass in scattering channel \\
\hline\hline
\end{tabular}
\end{center}
\caption{\small Mass parameters used in Eq.~(\ref{Opmodel}).}
\label{masses}
\end{table}

Before performing the calculations to all orders,
we write the transition operator as
\begin{eqnarray}
T & = & \left(\; 1\; -\; VG_{f}\;\right)^{-1}\; V
\label{Titeration} \\ & = &
\left\{\; 1\; +\; VG_{f}\; +\; VG_{f}VG_{f}\; +\; VG_{f}VG_{f}VG_{f}\; +\;
\cdots\;\right\}\; V
\; =\;
\sum_{N=0}^{\infty}\;\left( VG_{f}\right)^{N}\; V
\;\;\; ,
\nonumber
\end{eqnarray}

\noindent
where $G_{f}$ represents the free boson propagator
associated with the Hamiltonian $H_{f}$ in Eq.~(\ref{Opmodel}),
given by

\begin{equation}
G_{f}\left({{\vec{k}\,}'\;},{\vec{k}\;}; z\right)\; =\;
\braket{{{\vec{k}\,}'\;}}{\left( z-H_{f}\right)^{-1}}{\vec{k}\;}\; =\;
\fnd{2\mu}{2\mu z-k^{2}}\;
\delta^{(3)}\left(\vec{k}\, -{\vec{k}\,}'\right)
\;\;\; .
\label{Greensfu}
\end{equation}
The matrix elements of the $T$ operator follow from Eq.~(\ref{Titeration})
by

\begin{equation}
\bra{\vec{p}\,}\; T\;\ket{{\vec{p}\,}'}\; =\;
\int d^{3}k\;\sum_{N=0}^{\infty}\;
\bra{\vec{p}\,}\;\left( VG_{f}\right)^{N}\; \;\ket{{\vec{k}\,}}\;
\bra{\vec{k}\,}\; V\;\ket{{\vec{p}\,}'}
\;\;\; .
\label{pTpp}
\end{equation}
For the generalized potential~(\ref{generalpot}), the Born term
takes the form

\begin{equation}
V\left({\vec{p}\;},{\vec{p}\,}'\right)\; =\;
\bra{\vec{p}\,}\; V\;\ket{{\vec{p}\,}'}\; =\;
\bra{\vec{p}\,}\; V_{t}\;
\left( E(p)-H_{c}\right)^{-1}\; V_{t}\;
\ket{{\vec{p}\,}'}
\;\;\; .
\label{MSgeneralpot}
\end{equation}

\noindent
The total center-of-mass energy $E$ and the linear momentum $p$ are, through
Eq.~(\ref{Opmodel}), related by
\begin{equation}
E(p)\; =\;
\fnd{{\vec{p}\;}^{2}}{2\mu}\; +\; M_{1}\; +\; M_{2}
\;\;\; .
\label{Ep}
\end{equation}

\noindent
We denote the properly normalized eigensolutions of the operator $H_{c}$ in
Eq.~(\ref{Opmodel}), corresponding to the energy eigenvalue $E_{n\ell_{c}}$, by
\begin{equation}
\bracket{\vec{r}\,}{n\ell_{c}m_{c}}\; =\;
Y_{\ell_{c}m_{c}}\left(\hat{r}\right)\;
{\cal F}_{n\ell_{c}}(r)
\;\;\; ,\;\;\;
\xrm{with}\;\;\;
\left\{
\begin{array}{l}
\xrm{$n=0$, $1$, $2$, $\dots$}\\
\xrm{$\ell_{c}=0$, $1$, $2$, $\dots$}\\
\xrm{$m=-\ell_{c}$, $\dots$, $+\ell_{c}$}
\end{array}
\right.
\;\;\;\;.
\label{Vccomplete}
\end{equation}
So, by letting the self-adjoint operator $H_{c}$ act to the left in
Eq.~(\ref{MSgeneralpot}), we find
\begin{eqnarray}
\bra{\vec{p}\,}\; V\;\ket{{\vec{p}\,}'} & = &
\sum_{n\ell_{c}m_{c}}\;\bra{\vec{p}\,}\;
V_{t}\;\ket{n\ell_{c}m_{c}}\;\bra{n\ell_{c}m_{c}}
\;\left( E(p)-H_{c}\right)^{-1}\; V_{t}\;
\ket{{\vec{p}\,}'}
\nonumber\\ [.2cm] & = &
\sum_{n\ell_{c}m_{c}}\;\bra{\vec{p}\,}\; V_{t}\;
\fnd{\ket{n\ell_{c}m_{c}}\;\bra{n\ell_{c}m_{c}}}{E(p)-E_{n\ell_{c}}}
\; V_{t}\;\ket{{\vec{p}\,}'}
\;\;\; .
\label{MSgeneralpot1}
\end{eqnarray}
The individual matrix elements in the sum of formula~(\ref{MSgeneralpot1})
result in
\begin{equation}
\bra{n\ell_{c}m_{c}}\; V_{t}\;\ket{\vec{k}\,}\; =\;
\lambda\; (i)^{\ell}\;{\cal J}_{n\ell}(k)\;
Y^{\ast}_{\ell m}\left(\hat{k}\right)
\;\;\; ,
\label{nlmVk}
\end{equation}
where, in order to simplify the expressions, we define the quantities
\begin{eqnarray}
{\cal J}_{n\ell}(p) & = & \sqrt{\fnd{2}{\pi}}\;
\int r^{2}dr\; j_{\ell}\left( pr\right)\; U\left( r\right)\;
{\cal F}^{\ast}_{n\ell_{c}}\left( r\right)
\nonumber \\[2mm]
{\cal N}_{n\ell}(p) & = & \sqrt{\fnd{2}{\pi}}\;
\int r^{2}dr\; n_{\ell}\left( pr\right)\; U\left( r\right)\;
{\cal F}^{\ast}_{n\ell_{c}}\left( r\right)
\\ [2mm]
{\cal H}^{(1,2)}_{n\ell}(p) & = &
{\cal J}_{n\ell}(p)\;\pm i\; {\cal N}_{n\ell}(p)\; =\;
\sqrt{\fnd{2}{\pi}}\;
\int r^{2}dr\; h^{(1,2)}_{\ell}\left( pr\right)\; U\left( r\right)\;
{\cal F}^{\ast}_{n\ell_{c}}\left( r\right)
\;\;\; .
\nonumber
\label{calJcalH}
\end{eqnarray}

At this stage we should pay some attention to the quantum numbers of the
two coupled systems, since the operator $V_{t}$ mediates between systems
with different quantum numbers.
From parity conservation we deduce that the relative orbital angular
momenta $\ell$ of a meson-meson pair and $\ell_{c}$ of the confined
quark-antiquark pair differ by at least one unit.
Also the total intrinsic spins are different for the two systems, when
$C$ parity is taken into account.
The conservation of $J$, $P$, $C$, isospin, flavor, and color can be
formulated in a consistent way \cite{ZPC21p291,ZPC17p135,HEPPH0305056},
but we do not intend to go into details here.
We just should recall that there exists a relation between
($\ell_{c}$, $m_{c}$) and ($\ell$, $m$),
and moreover that the overlap integral expressing this relation
is absorbed into $\lambda$.

For the full Born term (\ref{MSgeneralpot1}), we then obtain
\begin{eqnarray}
\bra{\vec{p}\,}\; V\;\ket{{\vec{p}\,}'} & = &
\fnd{\lambda^{2}}{4\pi}\;\sum_{\ell}\; (2\ell +1)
P_{\ell}\left(\hat{p}\cdot{\hat{p}\,}'\right)\;
\sum_{n}\;
\fnd{{\cal J}^{\ast}_{n\ell}(p)\; {\cal J}_{n\ell}(p')}{E(p)-E_{n\ell_{c}}}
\;\;\; .
\label{MSgeneralpot2}
\end{eqnarray}

Similarly, one may determine the $N\!+\!1$-st-order term
$\bra{\vec{p}\,}\;\left( VG_{f}\right)^{N+1}\;\ket{{\vec{k}\,}}$
of Eq.~(\ref{pTpp}), yielding
\begin{eqnarray}
& &
\bra{\vec{p}\,}\;\left( VG_{f}\right)^{N+1}\;\ket{{\vec{p}\,}'}=
\fnd{\lambda^{2}}{4\pi}\;
\sum_{\ell}\; (2\ell +1)\;
\left[
\left( -i\;\pi\lambda^{2}\mu p\right)\;
\sum_{n}\;
\fnd{{\cal J}^{\ast}_{n\ell}(p)\; {\cal H}^{(1)}_{n\ell}(p)}
{E(p)-E_{n\ell_{c}}}
\right]^{N}\;
\times\nonumber\\ [.3cm] & & \;\;\;\;\times\;
P_{\ell}\left(\hat{p}\cdot{\hat{p}\,}'\right)\;
\sum_{n'}\;
\fnd{{\cal J}^{\ast}_{n'\ell}(p)\; {\cal J}_{n'\ell}(p')}
{E(p)-E_{n'\ell_{c}}}\;
\fnd{2\mu}{2\mu_{f} z-{p'}^{2}}\;
\;\;\; .
\label{termNp1}
\end{eqnarray}
From Eqs.~(\ref{MSgeneralpot2}) and (\ref{termNp1}), we find
for the matrix elements of the product $\left( VG_{f}\right)^{N}\; V$
the expression
\begin{eqnarray}
\lefteqn{\int d^{3}k\;
\bra{\vec{p}\,}\;\left( VG_{f}\right)^{N}\;\ket{{\vec{k}\,}}\;
\bra{\vec{k}\,}\; V\;\ket{{\vec{p}\,}'}\; =}
\nonumber\\ [.3cm] & &
\fnd{\lambda^{2}}{4\pi}\;
\sum_{\ell}\; (2\ell +1)\;
\left[
\left( -i\;\pi\lambda^{2}\mu p\right)\;
\sum_{n}\;
\fnd{{\cal J}^{\ast}_{n\ell}(p)\; {\cal H}^{(1)}_{n\ell}(p)}
{E(p)-E_{n\ell_{c}}}
\right]^{N}\;
\times\nonumber\\ [.3cm] & & \;\;\;\;\times\;
P_{\ell}\left(\hat{p}\cdot{\hat{p}\,}'\right)\;
\sum_{n'}\;
\fnd{{\cal J}^{\ast}_{n'\ell}(p)\; {\cal J}_{n'\ell}(p')}
{E(p)-E_{n'\ell_{c}}}\;
\;\;\; .
\label{TN}
\end{eqnarray}
The terms labelled with $N$ in Eq.~(\ref{TN}) can be summed up, resulting in
a compact final formula for the matrix elements of the transition
operator $T$ in Eq.~(\ref{pTpp}), reading

\begin{eqnarray}
\bra{\vec{p}\,} T\ket{{\vec{p}\,}'} & = &
\fnd{\lambda^{2}}{4\pi}\;
\sum_{\ell =0}^{\infty}(2\ell +1)\;
P_{\ell}\left(\hat{p}\cdot{\hat{p}\,}'\right)\;
\fnd{\dissum{n=0}{\infty}
\fnd{{\cal J}^{\ast}_{n\ell}(p){\cal J}_{n\ell}(p')}
{E(p)-E_{n\ell_{c}}}}
{1+i\pi\lambda^{2}\mu p
\left(\fnd{\lambda a}{\mu_{c}}\right)^{2}
\dissum{n=0}{\infty}
\fnd{{\cal J}^{\ast}_{n\ell}(p){\cal H}^{(1)}_{n\ell}(p)}
{E(p)-E_{n\ell_{c}}}} \;\;\;\; .
\nonumber\\ & &
\label{T}
\end{eqnarray}
\vspace{0.3cm}

{\bf Acknowledgments}:
We wish to express here our gratitude to Joseph Schechter
and his collaborators for their encouraging support to our work,
which has been very much appreciated by us, and
received as a good stimulus to continue our research.
One of us (EvB) wishes to thank the organizers of the Joefest conference
for kindly inviting him and for their warm hospitality.
He, moreover, thanks Christoph Hanhart, Francesco Sannino, and Jos\'{e}
Pel\'{a}ez for interesting and useful discussions.
This work was partly supported by the
{\it Funda\c{c}\~{a}o para a Ci\^{e}ncia e a Tecnologia}
of the {\it Minist\'{e}rio da
Ci\^{e}ncia e da Tecnologia} \/of Portugal,
under contract number
POCTI/\-FNU/\-49555/\-2002.

\end{document}